\documentclass[aps,prc,onecolumn,groupedaddress,showpacs,showkeys]{revtex4}
\usepackage{graphicx}

\begin{document}

\title{Baryon stopping  in the Color Glass Condensate formalism: A phenomenological study}

\author{F.O. Dur\~aes$^\dag$, A.V. Giannini$^\ddag$,  V.P.  Gon\c{c}alves$^\S$ and F.S. Navarra$^\ddag$ }
\affiliation{\dag\ Curso de F\'{\i}sica, Escola de Engenharia, Universidade Presbiteriana Mackenzie\\
CEP 01302-907, S\~{a}o Paulo, Brazil}
\affiliation{\ddag\ Instituto de F\'{\i}sica, Universidade de S\~{a}o Paulo\\
C.P. 66318,  05315-970 S\~{a}o Paulo, SP, Brazil}
\affiliation{\S\ Instituto de F\'{\i}sica e Matem\'atica,  Universidade Federal de Pelotas\\
Caixa Postal 354, CEP 96010-900, Pelotas, RS, Brazil\\}

\begin{abstract}
The net-baryon production at forward rapidities is investigated considering the Color Glass Condensate formalism. We assume that at large energies the coherence of the projectile quarks is lost and  that the leading baryon production mechanism changes from recombination to independent fragmentation. The phenomenological implications for net-baryon production in $pp/pA/AA$ collisions are analysed and predictions for LHC energies are presented.
\end{abstract}
\maketitle
\vspace{1cm}

\section{Introduction}

\date{\today}

In high energy hadronic collisions, baryons are produced both in the central  and in the forward rapidity region.
In the first case baryons are produced together with antibaryons and the net baryon number (baryons minus antibaryons) is
small. In contrast, in the large rapidity region there are almost only baryons and no antibaryons.  These experimental facts suggest that
the forward baryons are produced from the valence quarks of the projectile, whereas low rapidity baryons are produced mainly from gluons
and sea quarks. How valence quarks are converted into forward (or leading) baryons remains to be clarified. In lower ($\sqrt{s} \simeq 20 - 100$
GeV) energies proton-proton collisions, leading baryon production can be well understood in terms of recombination of the three valence quarks after
the collision with the target \cite{dnw}  or, equally well, in terms of diquark fragmentation \cite{bere}. At higher energies new phenomena are expected to
affect forward baryon production. At high energies  and at large rapidities, baryon production requires the interaction of valence quark with a relatively large momentum fraction
($x_1$) of the projectile with low fractional momentum (small $x_2$) partons in the target. In the low $x$ regime the target is a dense system of
partons (predominantly gluons) which may form the
Color Glass Condensate
(CGC), a state of very high partonic densities in which the nonlinear effects of QCD change the parton distributions and hence the cross sections (For reviews see Ref. \cite{hdqcd}) . The CGC is characterized by a momentum scale  which marks the onset of nonlinear (or saturation) effects. This so called saturation scale, $Q_s$, grows
with the reaction energy.  In Ref. \cite{strik} it was conjectured that at increasing projectile energies the valence quarks  receive a transverse momentum kick of the order of $Q_s$  and hence above a certain energy the coherence of the projectile quarks is lost and  the leading baryon production mechanism changes from recombination to independent fragmentation.  In  this work we shall explore the phenomenological implications of this assumption for the leading baryon production in $pp/pA/AA$ collisions at LHC energies. Our goal is to improve the previous studies using the CGC formalism that have been performed in Refs. \cite{tani,tani_prc},  where the nonlinear evolution of the target was accounted
for. In particular, we would like to improve the calculation of Ref. \cite{tani_prc} by
computing the $p_T$ distribution of the produced  leading baryons, which was missing in that work. Furthermore we also improve the treatment of the nonlinear effects, considering  the forward dipole scattering amplitude proposed in Ref. \cite{buw}, which capture the main properties of the solution of the BK equation, which determines the QCD evolution of the CGC, and describes the RHIC and LHC data for hadron production. We also extend these previous studies
 \cite{tani,tani_prc}  to $pp$ and $pA$ collisions and estimate for the first time the ratio   $R_{pA}  =
\frac{d^{2}N_{pA}}{dyd^{2}p_{T}}/A \frac{d^{2}N_{pp}}{dyd^{2}p_{T}}$ for leading baryon production. Finally, the proton and pion productions at forward rapidities are compared. Our study is strongly motivated by the recent results presented in Ref. \cite{mariovic}, which has demonstrated that the LHCf experimental data \cite{lhcf_pion} for the neutral pion production at very low-$p_T$ can be quite well described considering the CGC formalism, indicating the 
emergence of saturation scale as a hard momentum scale at very forward rapidities which allows to understand highly nonperturbative phenomena in QCD by using weak coupling methods.

This paper is organized as follows. In the next Section we present a brief review of the CGC formalism and its main formulas. In particular we present the models for the forward  dipole scattering amplitude used in our calculations.  In Section \ref{section:results} we present our results for the $p_T$ and $y$ dependences of the leading baryon cross section. A comparison with the RHIC data  is performed and predictions for baryon production in $pPb$ and $PbPb$ collisions at LHC energies are presented. Moreover, we present our predictions for the ratio  $R_{pA}$. Finally, in Section \ref{section:sum}, we summarize our main conclusions.

\section{Net baryon production in the CGC formalism}
\label{section:formulario}

In the CGC formalism the differential cross section for the forward production of a hadron of transverse momentum $p_T$ at rapidity $y$ reads \cite{jamaldumitruprl,kkt,dhj}
\begin{equation}
\label{dist}
\frac{dN}{d^2p_Tdy}= \frac{1}{(2\pi)^2 } \int_{x_{\text{F}}}^1 \frac{dz}{z^2} D(z)\frac{1}{q_T^2}\;x_1q_v(x_1)\;\varphi\left(x_2, q_T \right)\,,
\end{equation}
where the net-baryon fragmentation function is defined as:
\begin{equation}
D(z) \equiv D_{\Delta B/q}(z)=D_{B/q}(z)-D_{\bar B/q}(z)
\label{netfrag}
\end{equation}
with  $z=E_B/E_q$ being the  fraction of the energy of the fragmenting quark ($E_q$) taken by the emerging baryon $B$. The fractional momenta of the projectile quark and of the target gluon are $x_1=q_{T} \, e^{y}/\sqrt{s}$ and
$x_2=q_{T} \, e^{-y}/\sqrt{s}$ respectively.
The variable $q_T=\sqrt{p_T^2+m^2}/z$ is the quark transverse momentum and  the Feynman $x$ variable is given by
$x_F = \sqrt{p_{T}^2+m^2} \, e^{y}/\sqrt{s}$. Moreover,  $x_1 \, q_v(x_1)$ is the valence quark distribution of the projectile hadron and the function $\varphi(x_{2},q_{T})$ is the unintegrated gluon distribution of the hadron target which is given by:
\begin{equation}
\label{ugd}
\varphi(x_{2},q_{T}) = 2\pi q_{T}^{2} \int {r_{T}dr_{T} \mathcal{N}(x_{2},r_{T}) J_{0}(r_{T}q_{T})} \,,
\end{equation}
where $J_0$ is a Bessel function and $\mathcal{N}(x_{2},r_{T})$ is the forward scattering amplitude of a
color dipole of radius $r_T$ off a hadron target.

The evolution of $\mathcal{N}(x_{2},r_{T})$ is described in the mean field approximation of the CGC formalism \cite{CGC} by the BK equation \cite{bk}. This quantity  encodes the information about the hadronic scattering  and then about the non-linear and quantum effects in the hadron wave function (For reviews, see e.g. \cite{hdqcd}). In the last years, several groups have constructed phenomenological models which satisfy the asymptotic behaviours of the leading order BK equation in order to fit the HERA and RHIC data \cite{GBW,dipolos2,kkt,dhj,buw}. In general, it is  assumed that it can be modelled  through a simple Glauber-like formula, which reads
\begin{eqnarray}
{\cal{N}}(x,r_T) = 1 - \exp\left[ -\frac{1}{4} (r_T^2 Q_s^2)^{\gamma (x,r_T^2)} \right] \,\,,
\label{ngeral}
\end{eqnarray}
where $\gamma$ is the anomalous dimension of the target gluon distribution. The main difference among the distinct phenomenological models comes from the behaviour predicted for the anomalous dimension, which determines  the  transition from the nonlinear to the extended geometric scaling regime, as well as from the extended geometric scaling to the DGLAP regime. In this paper we restrict our analyses to the model proposed in Ref. \cite{buw}, the so called BUW model, which is able to describe  the $ep$ HERA data for the proton structure function and the hadron spectra measured in $pp$ and $dAu$ collisions at
RHIC energies \cite{buw,marcos_vic}.  Another feature of the BUW model which motivates this analysis is that it explicitly satisfies the property of geometric scaling \cite{scaling}, which is predicted for the solutions of the BK equation in the asymptotic regime of large energies. In the BUW model, the anomalous dimension is given by   $\gamma(x,r_T)=\gamma_s+\Delta\gamma(x,r_T)$, where  $\gamma_s = 0.628$ and  \cite{buw}
\begin{eqnarray}
\label{BUWeq}
\Delta \gamma(x,r_T) = \Delta \gamma_{BUW} =(1-\gamma_s)\frac{(\omega^a-1)}{(\omega^a-1)+b}.
\end{eqnarray}
In the expression above, $\omega  \equiv 1/(r_TQ_s(x))$ and the two free parameters $a=2.82$ and $b=168$ are fitted in such a way to describe the RHIC data on hadron production. It is clear, from Eq.(\ref{BUWeq}), that this model satisfies the property of geometric scaling \cite{scaling,marquet,prl}, since $\Delta\gamma$ depends on $x$ and $r_T$ only through the variable $1/r_TQ_s(x)$.
 Besides, in comparison with other phenomenological parametrizations, in the BUW model, the behaviour expected for the unintegrated gluon distribution in
the large $p_T$ limit (linear regime) is recovered: $\varphi(x_{2},q_{T})  \propto {1}/{q_T^4}$ at large $q_T$. In contrast, in Ref. \cite{tani} the nonlinear effects were taken into account considering the model proposed long ago by Golec-Biernat and Wusthoff \cite{GBW}, where the forward dipole scattering amplitude is given by Eq. (\ref{ngeral}) with $\gamma = 1$. This model implies that  the $r_T$ integration in Eq. (\ref{ugd}) can be carried out analytically and a simple expression for the
unintegrated gluon distribution can be obtained:
\begin{equation}
\label{ugd.anal}
\varphi(x_{2},q_{T}) = 4\pi \frac{q_{T}^{2}}{Q_{s}^{2}(x_{2})}\exp\left({-\frac{q_{T}^{2}}{Q_{s}^{2}(x_{2})}}\right) \,.
\end{equation}
 Although this model satisfactorily describes  the nonlinear regime (small - $q_T$), it clearly does not contain the expected behaviour for large-$q_T$. Consequently, the resulting predictions are not valid at large values of the transverse momentum of the hadron. This explains the behaviour observed in Figs. 3 and 4 of the Ref. \cite{tani_prc} for the net-proton spectra.

\section{Results }
\label{section:results}

\begin{figure*}
\begin{center}
\includegraphics[width=7cm]{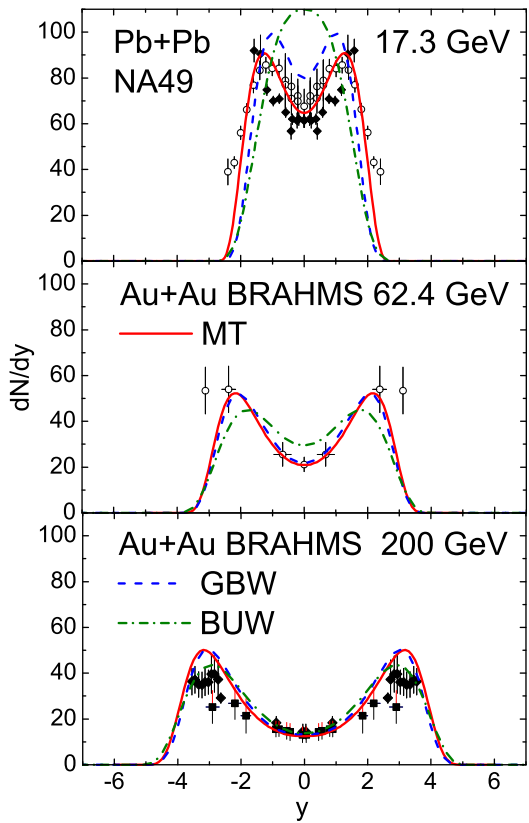}
\end{center}
\caption{(color online) Net-baryon rapidity distributions in  $PbPb$ collisions at  SPS energies and in $AuAu$ collisions at RHIC energies. Data from \cite{rhic,rhic2,rhic3,rhic4,rhic5,raploss}.}
\label{fig:1}
\end{figure*}

\begin{figure*}
\begin{center}
\includegraphics[width=6cm]{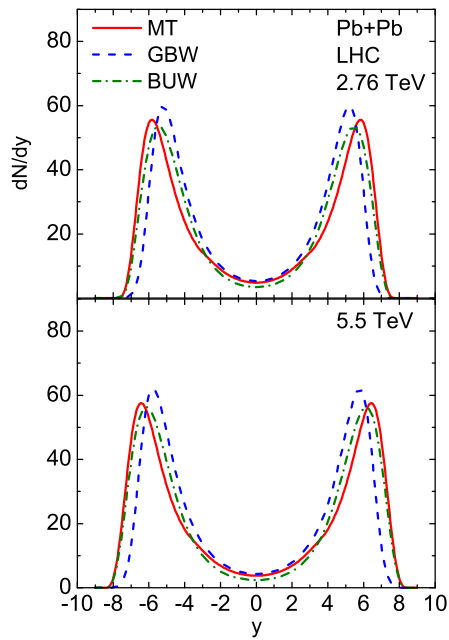}
\end{center}
\caption{(color online) Net-baryon rapidity distributions in central  $PbPb$ collisions at  LHC energies.  }
\label{fig:2}
\end{figure*}

\begin{figure*}
\begin{center}
\includegraphics[width=6cm]{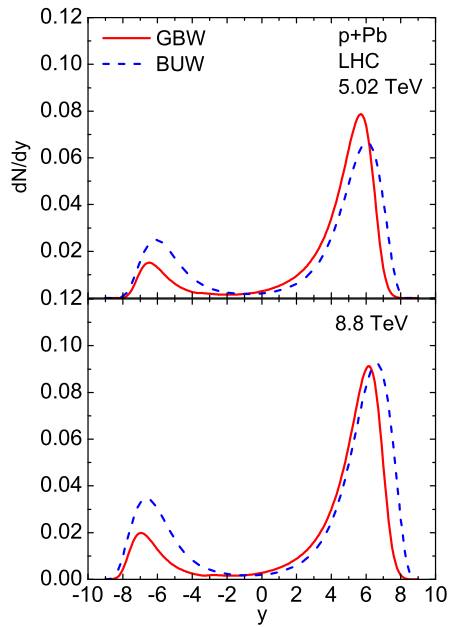}
\end{center}
\caption{(color online) Net-baryon rapidity distributions  in $pPb$ collisions at LHC energies.}
\label{fig:3}
\end{figure*}

Lets initially compare our predictions with the results previously presented  in Refs. \cite{tani,tani_prc}.
 In Fig.  \ref{fig:1} we present our results for the net-baryon rapidity distributions for
$PbPb$ collisions at SPS with energy $\sqrt{s} = 17.3$ GeV and also for $AuAu$ collisions at RHIC ($\sqrt{s} = 62.4$ and 200 GeV).
The net-baryon rapidity distribution is obtained integrating Eq. (\ref{dist}) in $p_T$  between $p_{T_{min}} = 0$  and
$p_{T_{max}} = \sqrt{s} \, e^{-y}$. The upper limit $p_{T_{max}}$ comes from the kinematical condition
$x_{F} < 1$. Following Ref. \cite{tani} we assume that the nuclear valence quark distribution is given by
$x \, q_v^{A}(x,Q^{2})= N_{\text{part}} x \, q_v^{proton}(x,Q^{2})$, with $N_{\text{part}}$ being the number of participants.
In Fig.  \ref{fig:1} the curve denoted MT represents the predictions originally derived in Ref. \cite{tani}, where the unintegrated gluon distribution
is given by the GBW model [Eq. (\ref{ugd.anal})], the proton valence quark distribution is described by the  MRST01-LO parametrization \cite{mrst01} and
the  fragmentation function is given by the following phenomenological model: $D_{p-\bar p}(z)=N\;z^a\;(1-z)^b$, with $N = 520142$, $a = 11.6$, $b = 6.74$. As already demonstrated in Refs. \cite{tani,tani_prc}, this model  describes quite well the experimental data for the net-baryon rapidity distribution for $\sqrt{s} = 17.4$ and 62.4 GeV, but overestimates the data for $\sqrt{s} = 200$ GeV and forward rapidities.
The GBW curve in Fig.  \ref{fig:1}  represents the predictions resulting from the substitution of the phenomenological model for the fragmentation function by the  KKP parametrization \cite{kkp}. We observe that the inclusion of a realistic model for fragmentation functions implies that the data at forward rapidities and very low energies are not well represented anymore. In contrast, at larger energies the MT and GBW predictions are similar. Finally,
the BUW curves show the predictions obtained considering the unintegrated gluon distribution derived from the BUW model for the forward dipole scattering amplitude. As in the GBW prediction, we also use  the  MRST01-LO and KKP parametrizations for the proton valence quark distribution  and   fragmentation functions, respectively. The BUW results improve the description of the data for  $\sqrt{s} = 200$ GeV and forward rapidities but fail to describe the data at smaller energies, which can indicate the limitation of this approach at lower energies. In Fig. \ref{fig:2} we present the BUW predictions for the net-baryon rapidity distributions in central $PbPb$ collisions at LHC energies. For comparison we also present the MT and GBW predictions. We observe that the three predictions are similar. For completeness, in Fig. \ref{fig:3} we present our predictions for $pPb$ collisions at LHC energies.

\begin{figure*}
\begin{center}
\includegraphics[width=12cm]{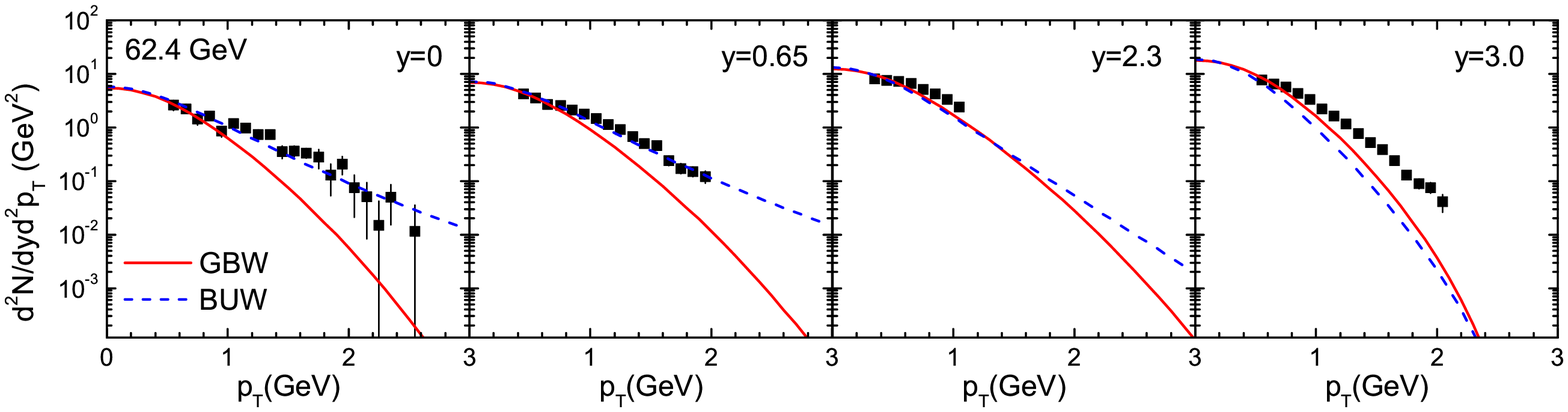}
\includegraphics[width=12cm]{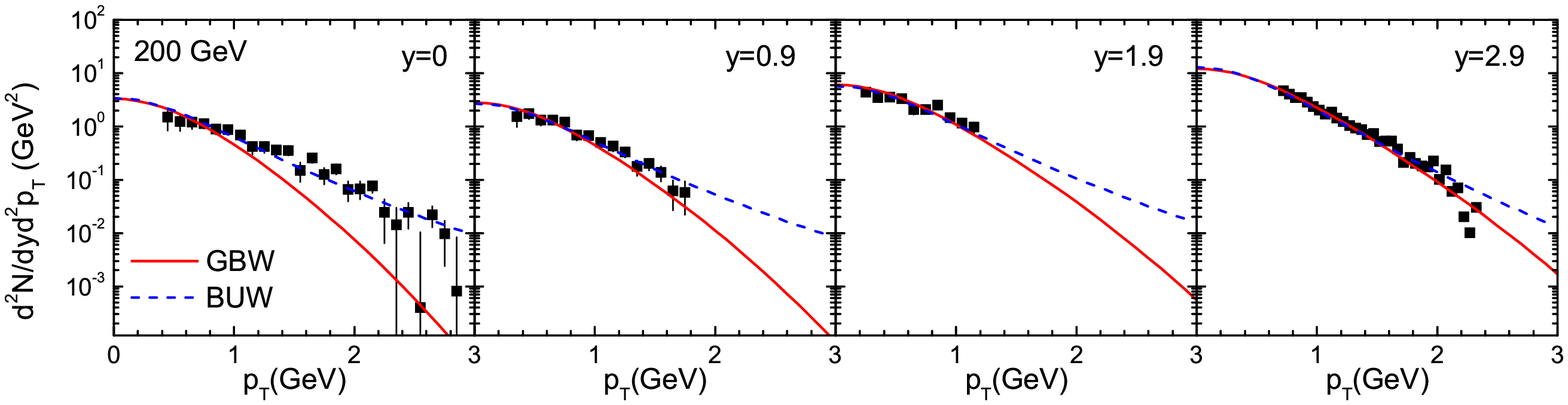}
\end{center}
\caption{(color online) Net-baryon transverse momentum spectra in central $AuAu$ collisions at  RHIC. Data from \cite{rhic,rhic2,rhic3,rhic4,rhic5,raploss}.}
\label{fig:4}
\end{figure*}

\begin{figure*}
\begin{center}
\includegraphics[width=12cm]{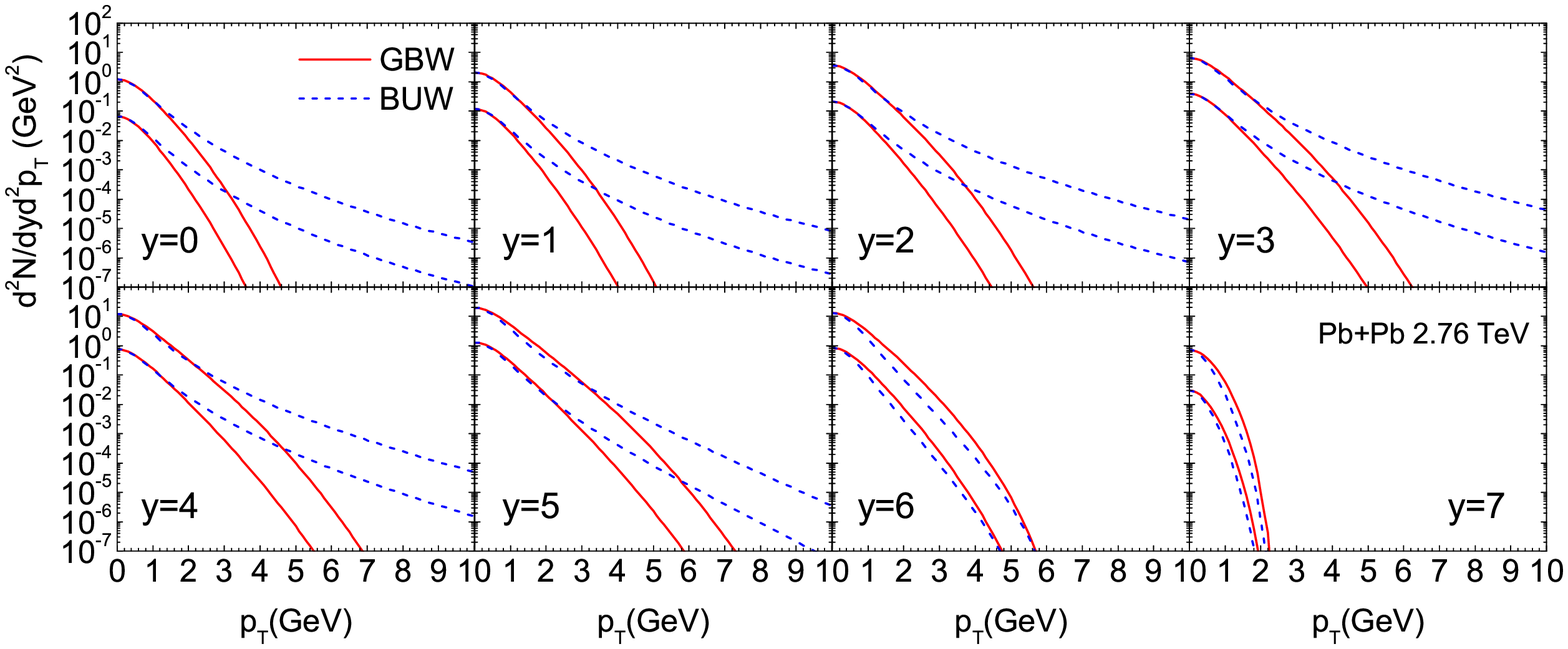}
\end{center}
\caption{(color online) Transverse momentum spectra  in $PbPb$ collisions at $\sqrt{s} = 2.76$ GeV and different rapidities. Upper and lower lines
represent pions and protons, respectively.}
\label{fig:5}
\end{figure*}

In Fig. \ref{fig:4} we present our predictions for the net-baryon transverse momentum spectra in central $AuAu$ collisions at  RHIC energies. As in Ref. \cite{tani_prc} we have assumed $N_{part}=  315$ and $357$ for  $\sqrt{s}=  62.4$ and $200$ GeV, respectively. These plots show two striking features. First, we observe a very good agreement between data and the spectra obtained with Eq. (\ref{dist}) and the BUW dipole amplitude and, at the same time, a disagreement between data points and the spectra obtained with the GBW dipole amplitude, specially when $p_T > 1$ GeV.
This happens because the GBW dipole amplitude has no DGLAP evolution and should not be able to reproduce data with large $p_T$.  The BUW amplitude
has the correct behaviour at larger $p_T$ and is able to describe the data in this region. Another interesting
feature of these plots is the failure of the formalism at the largest rapidity and lowest energy. This may be an indication that here the baryons are
not produced by independent quark fragmentation. They are more likely to be produced by coalescence of the incoming valence quarks. In Fig. \ref{fig:5} we present our predictions  for the proton and pion transverse momentum spectra at $\sqrt{s} = 2.76$ GeV and different rapidities. As already verified for RHIC energies, the GBW and BUW predictions are very distinct at large transverse momentum, in particular at $y \le 5$. At larger values of rapidities, both predictions are similar, which is directly associated to the limitation in the  phase space available for the considered energy. Moreover, it is important to emphasize the similarity between the $p_T$ behaviour of proton and pion production.

\begin{figure*}
\begin{center}
\includegraphics[width=12cm]{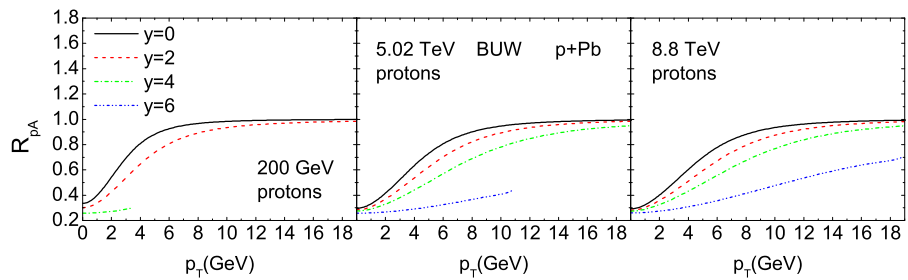}
\end{center}
\caption{(color online) Nuclear modification ratio, $R_{pA}$, for net-proton production in $pPb$ collisions at RHIC and LHC energies.}
\label{fig:6}
\end{figure*}

In order to quantify the magnitude of the nuclear effects in the net-proton production we introduce the   nuclear modification ratio, equal to the ratio of the net-proton production cross section in $pA$ collisions over the one in $pp$ collisions scaled by the number of binary collisions and defined by
\begin{equation}\label{RpA}
R_{pA} = \frac{\frac{d^{2}N_{pA}}{dyd^{2}p_{T}}}{A \frac{d^{2}N_{pp}}{dyd^{2}p_{T}}}\,\,.
\end{equation}
The behaviour of this ratio for  valence quark production, i.e. not including  quark fragmentation, has been studied in Ref. \cite{albakov} in the quasi-classical approximation of the McLerran-Venugopalan model \cite{MV} taking into account quantum corrections through the  nonlinear evolution derived in Ref. \cite{itakura}. The authors predict the presence of a Cronin enhancement in the quasi-classical regime and a suppression in the nuclear modification factor when the nonlinear effects are considered. In contrast to the approach discussed in Ref.  \cite{albakov}, which focus on the production of   soft valence
quarks far away (in rapidity) from the fragmentation region, here we consider the production of  hard valence quarks which experience no recoil and are
produced in the fragmentation region as proposed in Ref. \cite{jamaldumitruprl}. As already emphasized in Ref. \cite{albakov}, both approaches are complementary. However, the behaviour of $R_{pA}$ in the latter approach is still  an open issue. In Fig. \ref{fig:6} we present our predictions. We observe a suppression at small values of $p_T$ which increases at larger energies and rapidities, as expected from nonlinear effects.

\begin{figure*}
\begin{center}
\includegraphics[width=12cm]{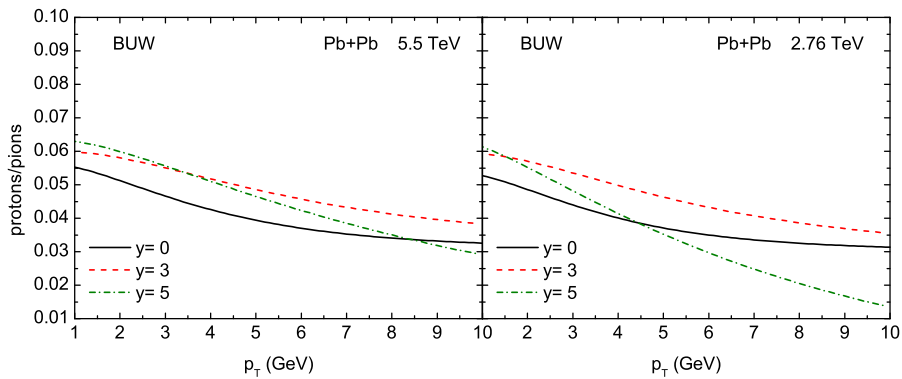}
\end{center}
\caption{(color online) Transverse momentum dependence  of the ratio between the cross sections for net-proton and net-pion production in $PbPb$
collisions at different values of rapidity and LHC energies.}
\label{fig:7}
\end{figure*}

The CGC formalism of forward particle production is appropriate to study the difference between the net-proton and net-pion production at forward rapidities. In what follows we analyse the behaviour of the ratio between the  cross sections for net-proton and net-pion production in $PbPb$ collisions. Until some years ago the proton to pion ratio was
expected to be always smaller than one. However, in some experiments \cite{ptopi-data},  this ratio was found to be much larger and in the
range $2 < p_T < 6$ GeV, being  even compatible with one. This has been called  "the baryon anomaly". Some explanations for this effect have been
proposed in \cite{coa,brod,topor}.  The interest in the subject will grow again now in view of the appearance of very recent data from ALICE \cite{chinelato}, which confirm the observation of the anomaly in $pPb$ collisions.
In Fig.\ref{fig:7} we show the proton-to-pion ratio as a function of the transverse momentum considering the CGC formalism.
As it can be seen, the $p/\pi$ ratio is small, depends very weakly on the rapidity, on the collision energy and  decreases with $p_T$. This is in sharp contrast with experimental data \cite{brod,topor,chinelato}, which show a ratio $p/\pi$ increasing with the
transverse momentum and reaching large values, close to $1$. Consequently, we conclude that in the CGC formalism  there is no baryon anomaly and pions are always more abundant. Therefore the anomaly must come from the protons and pions produced from gluons and sea quarks in the central rapidity region.

Forward nucleon production is very important for cosmic ray physics, where highly energetic protons reach the top of the atmosphere and undergo successive high energy
scatterings on the light nuclei in the air. In each of these  collisions, a projectile proton (the leading  baryon) looses energy, creating showers of particles, and goes
to the next scattering.  The interpretation of cosmic data depends on the accurate knowledge of the leading baryon momentum spectrum  and its energy dependence.
The crucial question of practical importance is the existence or non-existence of the Feynman scaling, which says that $x_F$-spectra of secondaries are energy independent.
In cosmic ray applications we are sensitive essentially to the large $x_F$ region (the fragmentation region) and hence  we can try to answer this question using the
CGC formalism and the expressions derived in the preceding sections. An additional motivation for this calculation is the fact that, in the near future,
Feynman scaling  (or its violation) will be investigated experimentally at the LHC by the LHCf Collaboration \cite{lhcf,lhcf2}.

\begin{figure*}
\begin{center}
\includegraphics[width=15cm]{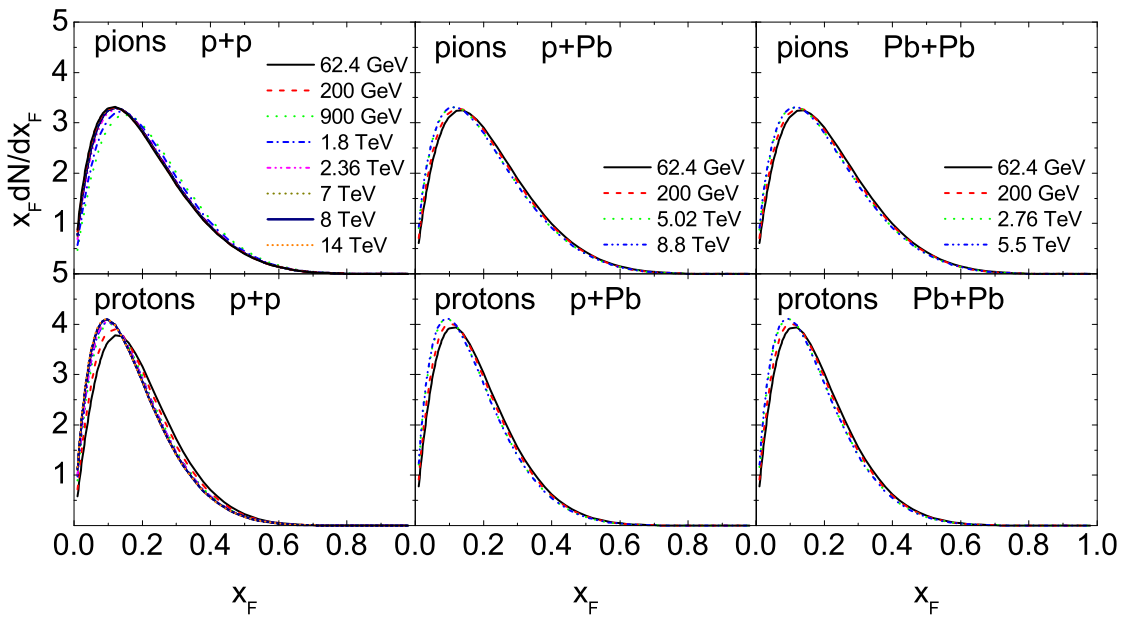}
\end{center}
\caption{(color online) Energy dependence of the $x_F$ distributions of pions and protons produced in $pp$, $pPb$ and $PbPb$ colisions.}
\label{fig.8}
\end{figure*}

Changing variables from $y$ to $x_F$ and integrating (\ref{dist}) over $p_T$ we obtain the  $x_F$ spectra  of leading protons and pions in $pp$, $pPb$ and $PbPb$
collisions at several energies, which are shown in Fig.  \ref{fig.8}.  In all panels we can clearly see a shift to smaller values of $x_F$, indicating a softening of
the leading particle spectrum. This Feynman scaling violation is compatible with the one obtained in Ref. \cite{dnw} and  Ref. \cite{merino}, where different mechanisms
are responsible for the violating behaviour.

\section{Conclusions}
\label{section:sum}

In this work we have improved the CGC formalism of forward particle production developed in  \cite{tani}, \cite{tani_prc},
\cite{jamaldumitruprl} and \cite{dhj} and applied it to the study of rapidity distributions, $p_T$  and $x_F$ spectra of
forward protons and pions.  We obtain a good agreement with existing data and show predictions for the forthcoming LHC data.

Concerning forward proton production, our results suggest that at energies around $\sqrt{s}=62.4$ GeV  there is a transition
from quark recombination to independent quark fragmentation. This is visible in Fig. \ref{fig:1}, where the independent fragmentation
dynamics underpredicts the data at large rapidities and lower energies but starts to describe the data very well at higher energies.
The same effect can be seen in Fig.  \ref{fig:4} at the largest rapidities.  A solid conclusion about this change of mechanism still
requires further theoretical and experimental work.   We observe a violation of Feynman scaling in  leading particle spectra which is
compatible with other approaches.  Finally, in the CGC formalism we do not observe any baryon anomaly. This suggests that this phenomenon
is related to the central region dynamics of gluons and sea quarks.

\section*{Acknowledgments}

This work was partially financed by the Brazilian funding agencies CAPES, CNPq,  FAPESP and FAPERGS.

\end{document}